\documentclass[12pt]{iopart}

\usepackage{threeparttable}
\usepackage{booktabs,multirow}
\usepackage{natbib}
\usepackage{setspace}

\usepackage{graphicx} 
\usepackage{flexisym}
\usepackage{textcomp,gensymb}
\usepackage{makecell}
\usepackage[T1]{fontenc} 

\newcolumntype{P}[1]{>{\centering\arraybackslash}p{#1}}

\begin{document}
\title[Flight model characterization of the telescope of the MATS satellite]{Flight model characterization of the wide-field off-axis telescope for the MATS satellite}

\author{Woojin Park$^1$, Arvid Hammar$^2$, Soojong Pak$^1,*$, Seunghyuk Chang$^3$, J\"{o}rg Gumbel$^4$, Linda Megner$^4$, Ole Martin Christensen$^4,5$, Jordan Rouse$^2$, and Dae Wook Kim$^6$}

\address{$^1$ School of Space Research and Institute of Natural Science, Kyung Hee University, Yongin 17104, Republic of Korea}
\address{$^2$ Omnisys Instruments AB, V\"{a}stra Fr\"{o}lunda SE-421 32, Sweden}
\address{$^3$ Center for Integrated Smart Sensors, Daejeon 34141, Republic of Korea}
\address{$^4$ Department of Meteorology (MISU), Stockholm University, Stockholm SE-106 91, Sweden}
\address{$^5$ Earth and Space Sciences, Chalmers University of Technology, Gothenburg SE-412 96, Sweden}
\address{$^6$ James C. Wyant College of Optical Sciences, University of Arizona, Tucson, Arizona 85721, USA}
\eads{\mailto{woojinpark@khu.ac.kr}, \mailto{soojong@khu.ac.kr, $^*$Corresponding author}}
\vspace{10pt}
\begin{indented}
\item[]July 2020
\end{indented}

\doublespacing

\begin{abstract}
We present optical characterization, calibration, and performance tests of the Mesospheric Airglow/Aerosol Tomography Spectroscopy (MATS) satellite, which for the first time for a satellite applies a linear-astigmatism-free confocal off-axis reflective optical design. Mechanical tolerances of the telescope were investigated using Monte-Carlo methods and single-element perturbations. The sensitivity analysis results indicate that tilt errors of the tertiary mirror and a surface RMS error of the secondary mirror mainly degrade optical performance. From the Monte-Carlo simulation, the tolerance limits were calculated to $\pm$0.5 mm, $\pm$1 mm, and $\pm$0.15$\degree$ for decenter, despace, and tilt, respectively. We performed characterization measurements and optical tests with the flight model of the satellite. Multi-channel relative pointing, total optical system throughput, and distortion of each channel were characterized for end-users. Optical performance was evaluated by measuring modulation transfer function (MTF) and point spread function (PSF). The final MTF performance is 0.25 MTF at 20 lp/mm for the ultraviolet channel (304.5 nm), and 0.25 - 0.54 MTF at 10 lp/mm for infrared channels. The salient fact of the PSF measurement of this system is that there is no noticeable linear astigmatism detected over wide field of view (5.67$\degree$ $\times$ 0.91$\degree$). All things considered, the design method showed great advantages in wide field of view observations with satellite-level optical performance.
\end{abstract}
\noindent{\it Keywords\/}: Optical telescopes (1174), Reflecting telescopes (1380), Wide-field telescopes (1800)

\maketitle

\section{Introduction}
\label{sec:intro}
\begin{figure}[!ht]
\centering\includegraphics[width=10cm]{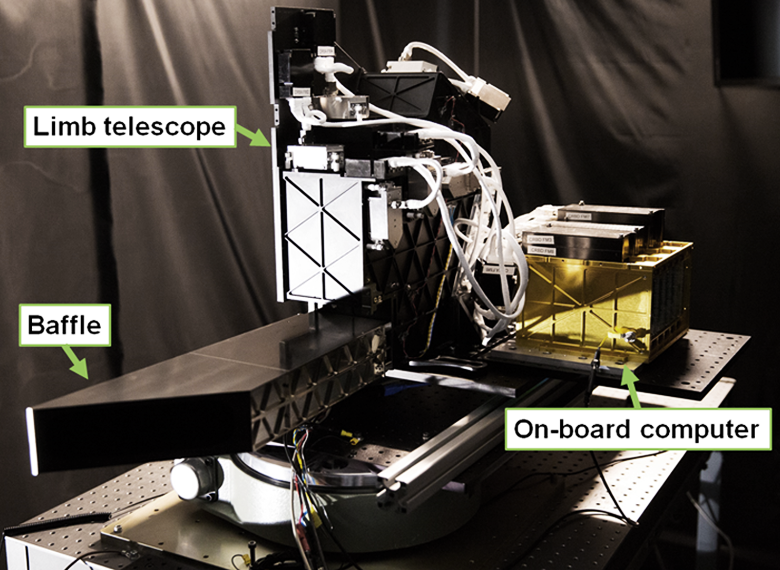}
\caption{\label{limb}A picture of the limb telescope with the baffle and the on-board computer.}
\end{figure}

Mesospheric Airglow/Aerosol Tomography Spectroscopy (MATS) is a Swedish microsatellite mission that observes noctilucent clouds (NLC) (80 - 86 km altitudes) and O\textsubscript{2} atmospheric band dayglow/nightglow (75 - 110 km altitudes) over wide field of view (5.67$\degree$ $\times$ 0.91$\degree$) in two ultraviolet (UV) channels, and four infrared (IR) channels within wavelength range between 270 - 772 nm \citep{jorg2020}. The main optical system of the MATS satellite is the limb-viewing telescope which is designed with a 35 mm entrance pupil diameter and a focal ratio of 7.4 \citep{hammar2018,hammar2019}. The whole system includes three off-axis mirrors, beam splitters, broad/narrow bandpass filters, and six Charge-Coupled Devices (CCDs) in the same compact limb housing (see Figure~\ref{limb}). Confocal off-axis reflective system is adapted to the telescope design for diffraction limited optical performance over full field of view. This optical design eliminates linear astigmatism without any correcting lenses, enabling wide field of view observations in a wide spectral range \citep{chang2004,chang2005, chang2006}. 

Tolerance analysis is generally performed before fabrication of the optical system to examine the feasibility of the optical design, but it is also required to confirm performance stability in satellite platform vibration environments \citep{wang2012}. Optical system characterization is also valuable for the final image corrections. Even though optical systems need to be calibrated during the mission \citep{cheng2018, grodecki2005}, laboratory characterization measurements before launching are essential \citep{wang2016, wang2017}. A distortion correction, especially, is crucial to increase accuracy of scientific results of the MATS satellite whose data use a tomography technique \citep{vo2019}. 

The linear-astigmatism-free (LAF) confocal off-axis reflective system has outstanding performance in low distortion and field curvature \citep{chang2015}. Performance of prototype LAF systems has been evaluated in previous research \citep{kim2010, parkLAF2020}, but characterization measurements have not been performed. Optical performance measurements for one of the IR channels of the prototype telescope has already been done \citep{hammar2019}. 

Since the limb telescope has six channels, performance and relative pointing accuracy for all channels must be carefully measured. All these tolerance analysis and characterization results can be utilized for the final imaging analysis. 

In this paper, sensitivity analysis and Monte-Carlo simulations performed for tolerance analysis are introduced in Section~~\ref{sec:tolerance}. Characterization of the telescope with total throughput, relative pointing, and distortion measurements of the system is introduced in Section~\ref{sec:calibration}. In Section~\ref{sec:imaging}, we show the imaging performance of the flight optics for two UV and four IR channels. All results are discussed and summarized in Section~\ref{sec:discussion}.
\section{Tolerance analysis}
\label{sec:tolerance}
Optical performance can be degraded by fabrication, assembly process, thermal conditions, vibration environments from the launch system, and satellite platform vibration environments, so tolerance analysis has been considered an important step in optical system development to improve reliability and practicality of the system \citep{wang2013, chen2015}. The main objectives of tolerance analysis are to determine optical performance degradation due to external environments and to decide tolerance ranges of the system for fabrication and alignment \citep{bauman2018}.

\begin{figure}[!ht]
\centering\includegraphics[width=10cm]{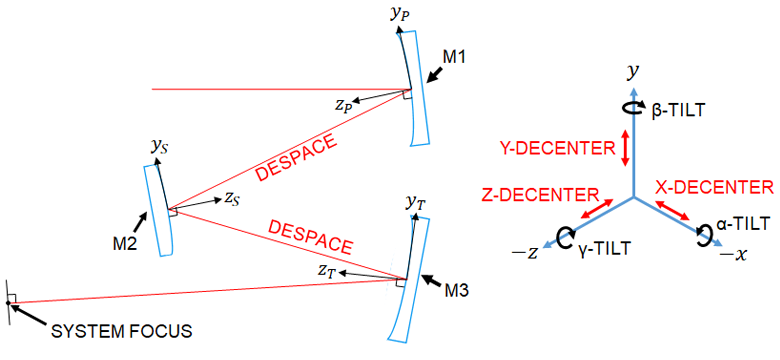}
\caption{\label{torcoord}The coordinate system for tolerance analysis.}
\end{figure}

Tolerancing parameters are x- and y- decenter, $\alpha$-, $\beta$-, and $\gamma$- tilt, despace, and focus. Decenter and tilt are adapted to each mirror, while despace corresponds to inter mirror distance \citep{kim2010}. The coordinate system for tolerance analysis is shown in Figure~\ref{torcoord}. Since we adjust the focal position to get the best image, the performance degradataion from tolerances is compensated by the focal position.

At the start, the individual tolerance budget of each parameter and their sensitivity are explored with sensitivity analysis. Then, a statistical analysis based on the Monte-Carlo method is performed to assess the system performance. Cumulative probability, which is the result of the Monte-Carlo simulation, enables us to estimate the expected final system performance.

\subsection{Sensitivity analysis}
\label{sec:sensitivity}
\begin{figure*}[t]
\centering\includegraphics[width=15.5cm]{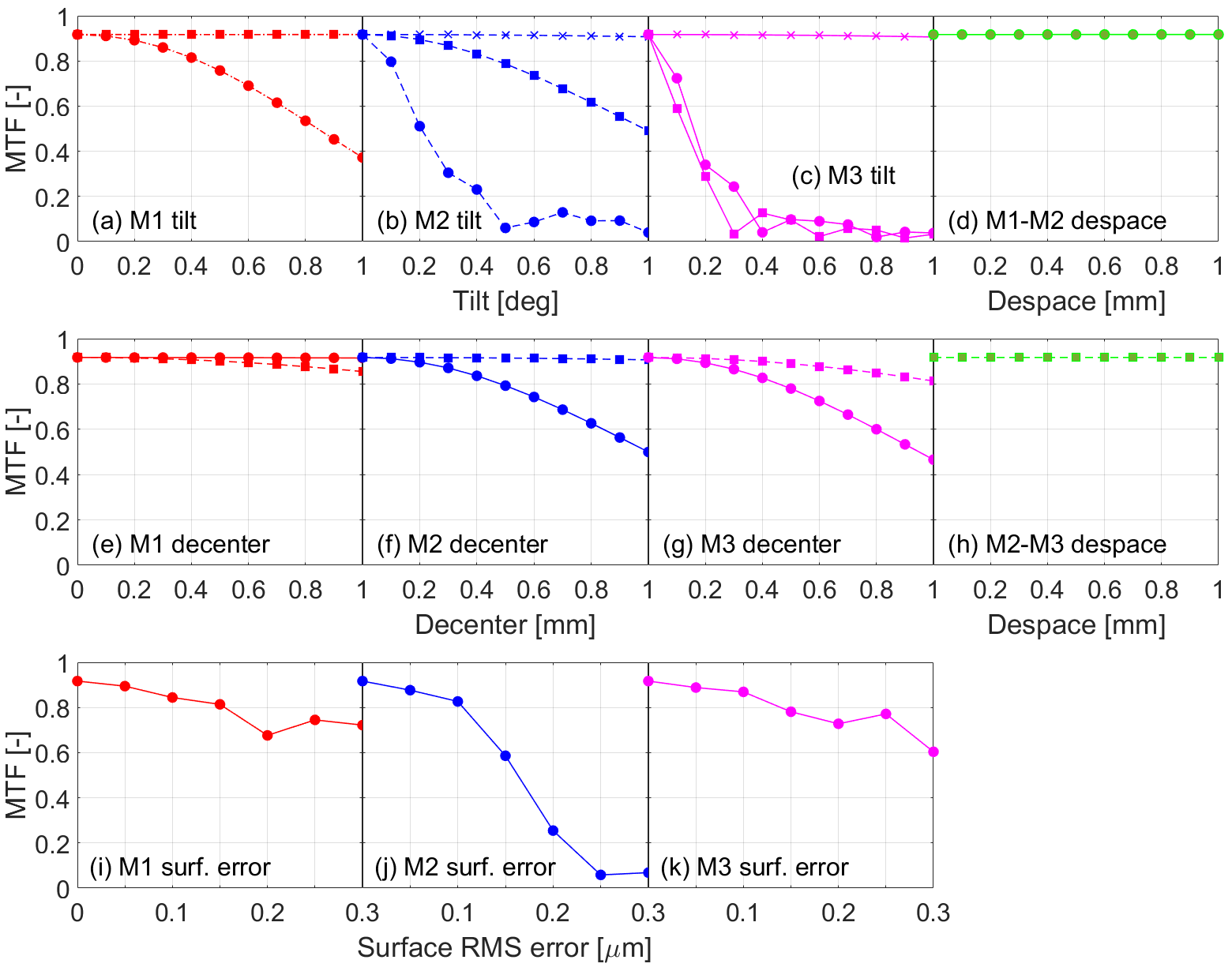}
\caption{\label{senstivitiy}Sensitivity analysis results of M1 (red), M2 (blue), and M3 (magenta): (a - c) $\alpha$- (circle), $\beta$- (square), and $\gamma$- (cross) tilt, (e - g) x- (circle) and y- (square) decenter, (i - k) surface RMS error, (d) M1-M2 (circle), and (h) M2-M3 (square) despace. $\gamma$-tilt of M1 is overlapped with its $\beta$-tilt.}
\end{figure*}

Optical component sensitivities are explored by individually implementing tolerance parameters \citep{lee2010}. The modulation transfer function (MTF) at 30 lp/mm is selected as the performance criterion, and the reference wavelength is set to 270 nm, which is the most sensitive channel to mirror surface errors. MTF values at five points within the field of view  (i.e., -2.84$\degree$ $\times$ -0.46$\degree$, -1.42$\degree$ $\times$ -0.23$\degree$, 0.00$\degree$ $\times$ 0.00$\degree$, 1.42$\degree$ $\times$ 0.23$\degree$, 2.84$\degree$ $\times$ 0.46$\degree$) are averaged for each tolerance parameter. Figure~\ref{senstivitiy} illustrates sensitivity analysis results of each tolerance parameter. All parameters of M1 (primary mirror), M2 (secondary mirror), and M3 (tertiary mirror) correspond to red, blue, and magenta colors for plots (a - c), (e - g), and (i - k). Inter mirror distances of M1-M2 and M2-M3 are indicated with green circles and green squares, respectively.

From the sensitivity analysis, we expect that $\alpha$- tilt of M2, $\alpha$- and $\beta$- tilt of M3, and surface RMS error of M2 are the parameters most critical for image quality degradation. On the other hand, the image quality is significantly less sensitive to $\gamma$- tilt and despace for all mirrors. In this system, $\alpha$- or $\beta$- tilt of M3 is critical for image quality, so it can be set to the compensator for realignments.

\subsection{Monte-Carlo simulation}
\label{sec:montecarlo}
Sensitivity analysis can provide performance sensitivity for each of the optical component errors. However, it is necessary to confirm the system tolerance limits when all tolerances simultaneously affect the system. The Monte-Carlo method is the most common way to predict the cumulative probability for meeting specific performance requirements  \citep{kus2017, burge2010}.

Initial tolerance limits are estimated from the root sum square of each parameter of three mirrors calculated from sensitivity analysis, and they are optimized within fabrication and alignment error budgets \citep{funck2010}. OpticStudio is used for the Monte-Carlo simulation. Table~\ref{table:torlimit} lists tolerance parameters and the final tolerance limits that are calculated using the iterative method. Tolerance limits are the same for all mirrors. Focus is selected as a compensator, and reference wavelength is the same as the one used in the sensitivity analysis. 

\begin{table}[!ht]
\centering
\caption{Tolerance limits for the Monte-Carlo simulation}
\footnotesize
\begin{threeparttable}
\label{table:torlimit}
\begin{tabular}{P{3cm}|P{5cm}}
\toprule
Parameter & Tolerance limits\tnote{a} \\\hline
x-, y- Decenter & $\pm$0.5 mm \\
$\alpha$-, $\beta$- Tilt & $\pm$0.15$\degree$\\
Despace & $\pm$1.0 mm \\
Focus\tnote{b} & $\pm$1.0 mm \\\bottomrule
\end{tabular}
\begin{tablenotes}
\item[a]{\footnotesize Tolerance limits are the same for M1, M2, and M3.}
\item[b]{\footnotesize Focus is used as the compensator.}
\end{tablenotes}
\end{threeparttable}
\end{table}
\vspace{3 mm}

\begin{figure}[!ht]
\centering\includegraphics[width=8cm]{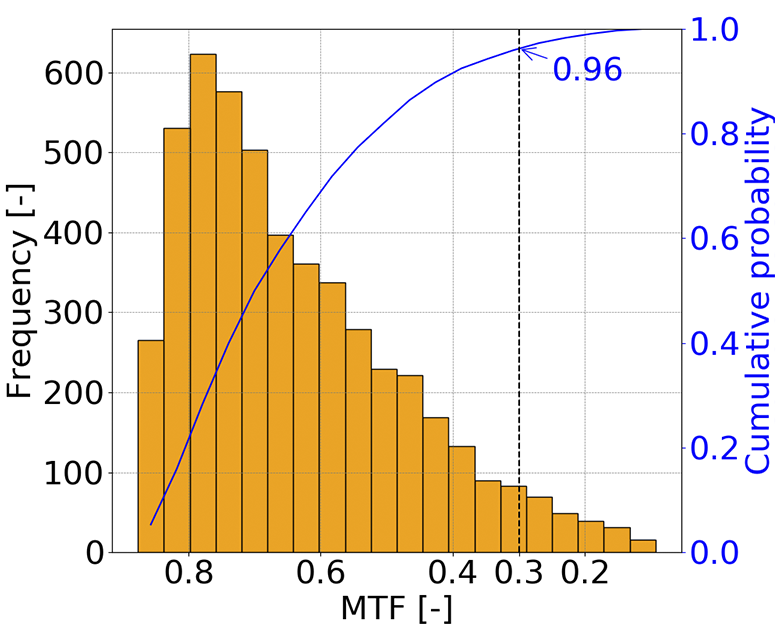}
\caption{\label{monte}Monte-Carlo simulation results. The performance limit is indicated by a black dashed line. The blue solid line represents a cumulative probability curve.}
\end{figure}

Tolerance distributions for the Monte-Carlo simulation follow a normal distribution. Figure~\ref{monte} shows the histogram of 5,000 Monte-Carlo tries that are binned as a function of MTF. Required optical performance (i.e., 0.3 MTF) is met at 96 $\%$ cumulative probability. From the sensitivity analysis of the surface RMS errors, we expect that $\sim$0.03 MTF could be additionally degraded when taking count the fabricated surface RMS errors, which are 0.049, 0.034, and 0.062 $\mu$m for M1, M2, and M3, respectively~\citep{hammar2019}.

Tolerance limits of x- and y- decenter and $\alpha$- and $\beta$- tilt are allowable ranges for the mission requirement when considering the mirror sizes that are 60 (L) mm $\times$ 40 (W) mm, 36 mm $\times$ 36 mm, and 90 mm $\times$ 80 mm for M1, M2, and M3, respectively. Nominal despace is $\sim$250 mm for both M1-M2 and M2-M3. The tolerance limit of despace is less strict since it rarely affects optical performance if focus compensations are mechanically available within $\pm$1.0 mm (see (d, h) in Figure~\ref{senstivitiy}).

Tilt and decenter errors can be compensated thanks to shims and L-brackets that are used to precisely position the mirrors and can be chosen in different thicknesses for relocations of the optical components.

\section{System calibration and characterization}
\label{sec:calibration}
Optical system calibration allows estimations of power and incident angles of the light entering the entrance pupil \citep{hagen2014}. It considers not only systematic noise corrections that are bias, dark, and flat-field corrections in CCD data \citep{birney2005}, but also characterization of total system throughput, relative pointing, distortion, and etc. Calibrations for electronics of the limb house have been performed \citep{giono2018}. Systematic noise will be subtracted (bias and dark) and divided (flat) from the raw object frame after the data acquisition. The limb telescope has six channels and 19 optical components including mirrors, beam splitters, broad, and narrow bandpass filters. It is important to characterize total system throughput and relative pointing of each channel.

\subsection{Total system throughput}
\begin{figure}[!ht]
\centering\includegraphics[width=10cm]{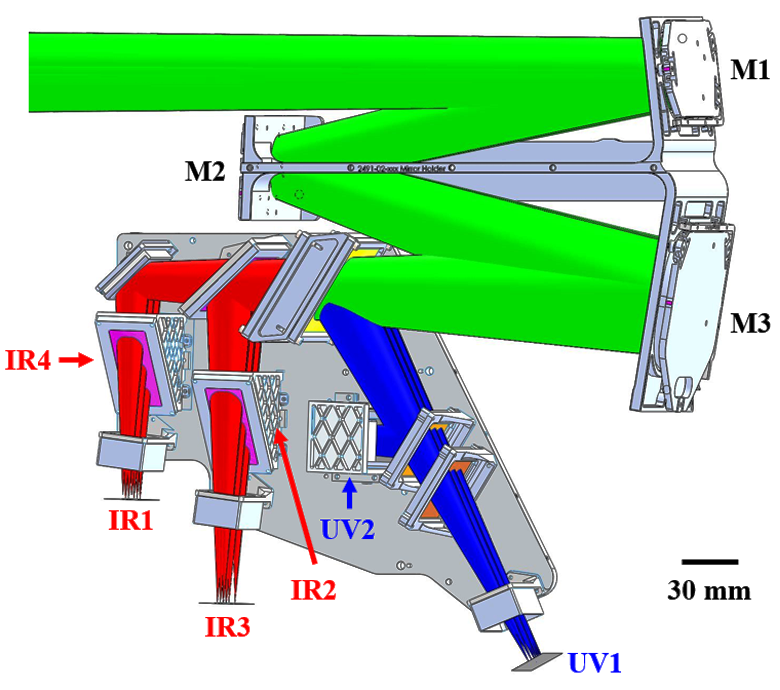}
\caption{\label{opticpath}The optical layout of the limb telescope. UV2, IR2, and IR4 CCDs are located at the backside of the instrument.}
\end{figure}

In the splitter box of the MATS telescope system, the incident beam from the off-axis telescope is split into six channels (see Figure~\ref{opticpath}). Verification of transmittance and reflectivity of each optical component is necessary to calculate total throughput of the system and then to decide the CCD gain, the exposure time, and etc. Furthermore, reflectivity of diamond turned aluminum mirrors needs to be measured for the characterization of the scattered light from high surface roughness mirrors \citep{harvey2013, ingers1989}.

Transmittance and reflectivity of filters and beam splitters have been measured by the component providers. We used the Andor iDus spectrometer and 1kW Xenon lamp to measure specular reflectivity of the mirrors. Measurement results are listed in Table~\ref{table:throughput}. Beam splitters, broadband filters, narrowband filters, and folding mirrors are abbreviated as BS, FB, FN, and FM, respectively. All the reflectivity and transmittance values are based on the central wavelength of each channel.

\begin{table*}[t]
\centering
\caption{Bandpass, reflectivity, and transmittance of optical components\tnote{a}.}
\footnotesize
\begin{threeparttable}
\label{table:throughput}
\begin{tabular}{P{1.2cm}|P{1.6cm}|P{1.20cm}|P{0.8cm}|P{0.8cm}|P{0.8cm}|P{0.8cm}|P{0.8cm}|P{0.8cm}|P{1.0cm}|P{0.8cm}}
\toprule
Channel & WL (nm)       & Mirror (-)             & BS1 (-)  & FB1 (-)   & BS2 (-)   & FM (-)    & BS3 (-)   & FN (-)    & CCD QE (-)    & Total ($\%$) \\\hline
UV1     & 270$\pm$1.5   & 0.86                  & 1.0 (R)  & 0.70      & 0.50 (T)  & -         & -         & 0.38      & 0.50          & 4.2 \\\hline
UV2     & 304.5$\pm$1.5 & 0.80                  & 1.0 (R)  & 0.69      & 0.51 (R)  & 0.79      & -         & 0.41      & 0.55          & 3.2 \\\hline
IR1     & 762$\pm$1.8   & 0.88                  & 0.96 (T)  & 0.97      & 0.45 (T)  & 0.88      & 0.60 (T)  & 1.0      & 0.78          & 12\\\hline
IR2     & 763$\pm$4.0   & 0.88                  & 0.96 (T)  & 0.97      & 0.55 (R)  & -         & 0.75 (R)  & 0.98      & 0.78          & 20\\\hline
IR3     & 754$\pm$1.5   & 0.88                  & 0.96 (T)  & 0.98      & 0.55 (R)  & -         & 0.25 (T)  & 0.77      & 0.80          & 5.4\\\hline
IR4     & 772$\pm$1.5   & 0.88                  & 0.96 (T)  & 0.98      & 0.45 (T)  & 0.88      & 0.40 (R)  & 0.81      & 0.76          & 6.2 \\\bottomrule
\end{tabular}
\begin{tablenotes}
\item[a]{\footnotesize WL: Wavelength, BS: Beam splitter, FB: Broadband filter, FN: Narrowband filter, FM: Folding mirror, QE: Quantum efficiency, (R): Reflectivity, (T): Transmittance}
\end{tablenotes}
\end{threeparttable}
\end{table*}
\vspace{3 mm}

IR1 and IR2 show relatively high total throughput ($>$ 12 $\%$) while other channels have throughput between 3.2 - 6.2 $\%$. Low throughput is already expected by design. However, the MATS satellite observes the earth mesosphere, which is bright enough with the adequate CCD gain, and exposure time ($\sim$3 seconds). The three off-axis mirrors have UV enhanced aluminum coating. They are supposed to have 89 $\%$ and 85 $\%$ reflectivity for the UV and IR ranges, respectively \citep{edmund}. The measurement results show that the mirror reflectivity in the UV channels is lower than our expectations, which might be the result of scattering caused by high surface roughness ($\sim$3 nm) \citep{schmitt1990}.

\subsection{Relative pointing characterization}
Total throughput measurements and characterization of optical aberrations are common optical calibration tasks for telescopes \citep{park2012}, but relative pointing measurements are also necessary for multi-channel telescopes. As we mentioned in Section~\ref{sec:intro}, two observation targets, noctilucent clouds and O\textsubscript{2} atmospheric band dayglow/nightglow, are located at different altitudes. For this reason, the image centers of UV and IR channels are different by design: The four IR channels should share their field of view, and so should the two UV channels.

\begin{figure}[!ht]
\centering\includegraphics[width=10cm]{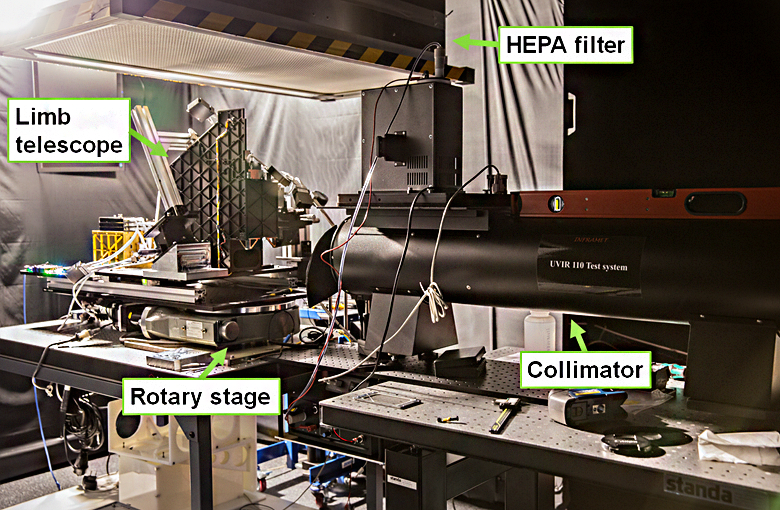}
\caption{\label{testsetup}The optical test setup for characterization and optical performance measurements. The limb telescope is installed under the HEPA filter and the fan (left-side), and the collimator sits in front of the limb entrance aperture (right-side).}
\end{figure}

There are relative pointing errors resulting from mechanical fabrication errors, filter or beam splitter misalignments, and etc. Relative pointing measurements enable end-users to know and correct the pointing errors of each channel so that the proper targets can be observed. 
The relative pointing between the channels was measured by taking point source images that were generated by the 100 $\mu$m pinhole and the Inframet CDT11100HR collimator (Figure~\ref{testsetup}) \citep{inframet}. This optical test setup was also used for imaging performance measurements (Section~\ref{sec:imaging}).

\begin{figure}[!ht]
\centering\includegraphics[width=10cm]{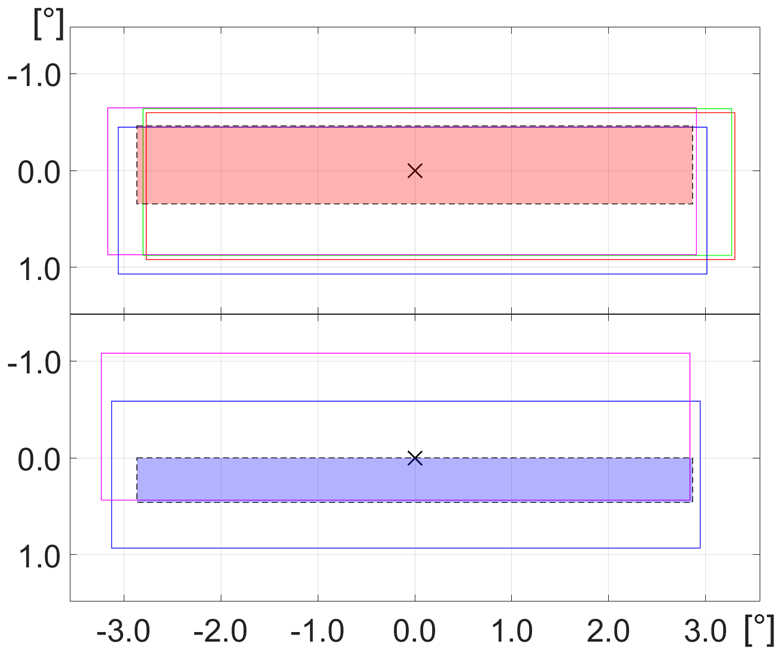}
\caption{\label{relativepointing}Relative pointing of IR (top) and UV (bottom) channels. Field of view of IR1 (red), IR2 (magenta), IR3 (green), IR4 (blue), UV1 (blue), and UV2 (magenta) are illustrated with solid lines.}
\end{figure}

Figure~\ref{relativepointing} displays the relative pointing of the IR and UV channels. Black crosses indicate the pointing reference of the satellite that is also considered as the optical axis of the telescope. Red and blue areas show targets for each channel. The field of view for each CCD is overlaid with colored solid lines. 

There are tight margins in the vertical direction, especially for UV channels. The IR channels seem to have good alignments while covering the target in all four channels. The maximum separations among IR channels are 0.2$\degree$ in vertical and 0.4$\degree$ in horizontal. UV channels are misaligned to the vertical direction by 0.5$\degree$ while they horizontally misaligned by 0.1$\degree$. Even though UV fields are largely separated, they still properly cover their target.

The observed images should be aligned by using relative pointing offset values listed in Table~\ref{table:relativepoint}. The pointing reference (0.00, 0.00) corresponds to black crosses in Figure~\ref{relativepointing}. Due to different altitudes of UV and IR targets (70 - 90 km for UV, and 75 - 110 km for IR), y-offsets of each source deviate from 0.00 pixels.

\begin{table}[!ht]
\centering
\caption{Relative pointing offsets of UV and IR targets. Offset values are relative to the pointing reference.}
\footnotesize
\begin{threeparttable}
\label{table:relativepoint}
\begin{tabular}{P{3cm}|P{2cm}|P{2cm}}
\toprule
Target / Channel  & x-offset (pix)\tnote{a}   & y-offset (pix) \\\hline
UV source       & 0.00                      & 77.60\\
UV1             & -30.72                    & 58.54\\
UV2             & -66.56                    & -108.38\\
IR source       & 0.00                      & -19.40\\
IR1             & 89.79                     & 54.47\\
IR2             & -44.02                    & 37.64\\
IR3             & 78.58                     & 40.31\\
IR4             & -7.49                     & 104.92\\\bottomrule
\end{tabular}
\begin{tablenotes}
\item[a]{\footnotesize The pixel size is 13.5 $\mu$m square that is for the E2V CCD42-10 CCD.}
\end{tablenotes}
\end{threeparttable}
\end{table}
\vspace{3 mm}

\subsection{Distortion}
The MATS satellite will generate 3D cube data by using the tomography technique, which combines a bunch of images. Each frame would not match together if distortion exists, and it creates large errors in the tomography. Distortion is measured with the point source and a distortion target \citep{inframet}. In this test, we accurately rotated the telescope into a specific angle, and compared the rotation angle with the incident angle of the beam derived from the image location at the sensor.
\begin{figure}[!ht]
\centering\includegraphics[width=10cm]{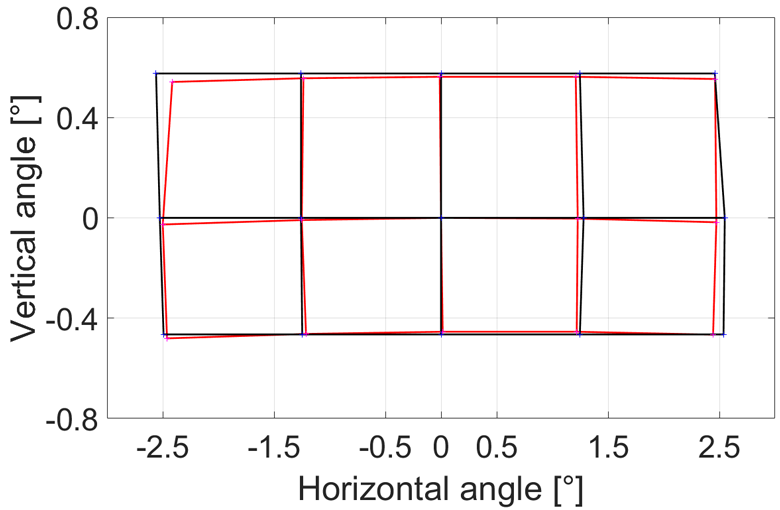}
\caption{\label{distortion}Distortion measurements of the limb telescope. Black solid lines indicate the distortion grid by design, while red solid lines represent the measured distortion grid.}
\end{figure}

As the result shown in Figure~\ref{distortion}, there is pincushion-like distortion with more aberrations to the upper-left and the lower-right corners than to the other corners. The maximum distortions to vertical and horizontal directions are 2.57 $\%$ and 3.70 $\%$, respectively. Distortions by design are 0.08 $\%$ and 2.83 $\%$ in vertical and horizontal directions.

Measured distortion can be generated by mirror surface figure errors that mainly come from the fabrication process and by filter bending cased by assembly stress.

\section{Imaging and MTF performance}
\label{sec:imaging}
Imaging performance measurements took place in an ISO class 5 cleanroom. The limb instrument was installed on a rotary stage so that the optical tests could be performed not only at the image center, but also over full image fields (Figure~\ref{limb} and \ref{testsetup}). Before we evaluated the optical performance, we precisely found focal positions by measuring 80 $\%$ encircled energy diameter (EED) of point source images. 

\begin{figure*}[!ht]
\centering\includegraphics[width=15cm]{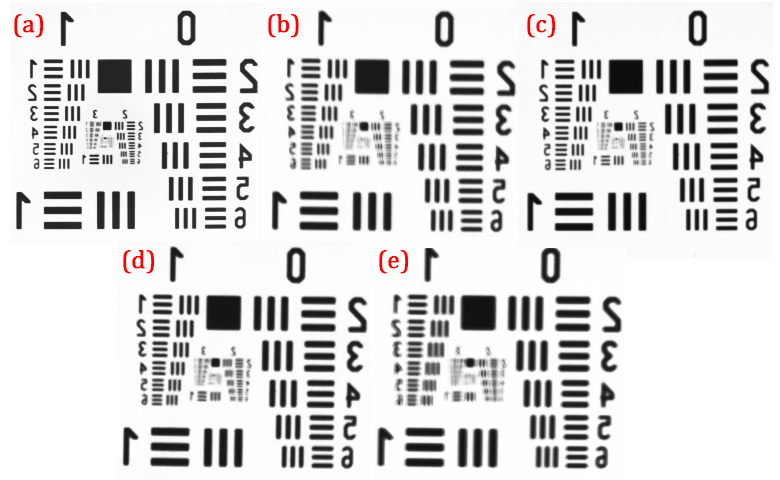}
\caption{\label{usaf}USAF target images from the limb telescope in (a) UV2, (b) IR1, (c) IR2, (d) IR3, and (e) IR4. A USAF image in UV1 not exists because of the low transmission of the target in 270$\pm$1.5 nm wavelength.}
\end{figure*}

The 1951 USAF target is a good indicator for finding the approximate focus and for visually inspecting the optical performance. Figure~\ref{usaf} shows the USAF target images in UV2, IR1, IR2, IR3, and IR4 displaying clear three individual bars at the group 0 element 4 and group 1 element 4 in all channels, fullfilling the requirements of the IR channels. For the UV channels, it is required that the lines can be separated for group 2 element 4. Looking at Figure~\ref{usaf} (a), one can see that this requirement is fulfilled for channel UV1. It was not possible to take a USAF target image in UV1 due to extremely low transmission of the target in a 270$\pm$1.5 nm wavelength band.

To determine the MTF, a slanted edge test target was used. Sharpness of the edge across the image position can be expressed with the edge spread function (ESF). It is transformed into line spread function (LSF) by taking derivative of the ESF. The final MTF curves are derived by taking the Fourier transform and normalizing it \citep{fujita1992, judy1976, padgett2006, samei1998, zhang2012}.

\begin{figure}[!ht]
\centering\includegraphics[width=10cm]{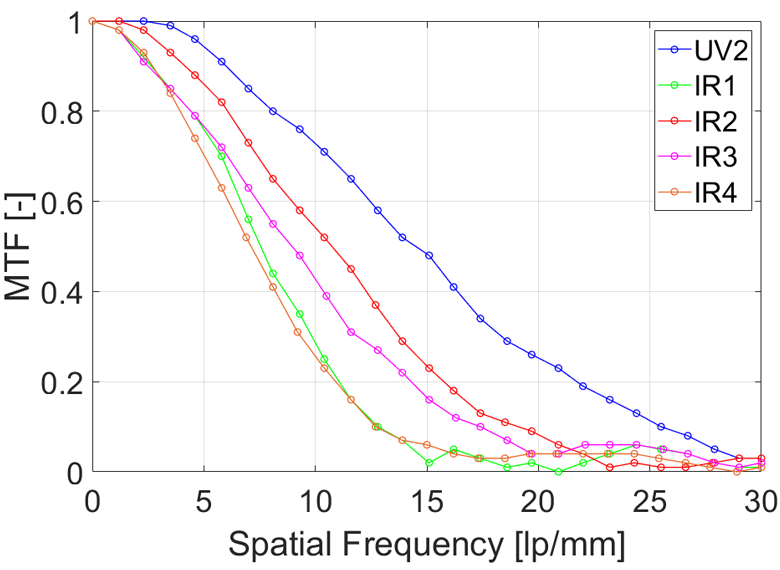}
\caption{\label{mtf}Measured MTF curves of UV and IR channels except for UV1.}
\end{figure}

The MTF measurements show that we have the best image resolution in UV2 followed by IR2, IR3, IR1, and IR4 (Figure~\ref{mtf}). Measured MTF performance is 0.25 MTF at 20 lp/mm for UV2, and 0.25, 0.29, 0.44 ,and 0.54 MTF at 10 lp/mm for IR4, IR1, IR3, and IR2, respectively. The MTF curve for UV1 was not available for measuring because of the low UV1 transmission to the slanted edge target.

Instead, we evaluated the performance in UV1 with a point source image by comparing it to that of other channels. Measured spot sizes are listed in Table~\ref{table:spotsize}. As we expected from the MTF chart, UV2 has the smallest spot size out of all the channels. The performances among the channels for MTF and 80 $\%$ EED follow the same trend except for IR1. The difference is caused by the big pixel size of the CCD (i.e., 13.5 $\mu$m), which implies that the pixel coverage of the point source image is only two to three pixels, making it difficult to estimate the exact spot size. UV1 has the biggest spot size among six channels. If we consider uncertainty from undersampling of 80 $\%$ EED, the actual optical performance for UV1 will approximately equal to that of IR4.

\begin{table}[!ht]
\centering
\footnotesize
\caption{80 $\%$ EED of 100 $\mu$m point source images}
\begin{threeparttable}
\label{table:spotsize}
\begin{tabular}{P{2cm}|P{2cm}|P{3cm}}
\toprule
Channel & WL (nm)       &80 $\%$ EED ($\mu$m) \\\hline
UV1 	& 270$\pm$1.5   &57.81 \\
UV2 	& 304.5$\pm$1.5 &26.72 \\
IR1 	& 762$\pm$1.8   &33.85 \\
IR2 	& 763$\pm$4.0   &38.19 \\
IR3 	& 754$\pm$1.5   &38.94 \\
IR4 	& 772$\pm$1.5   &48.22 \\\bottomrule
\end{tabular}
\end{threeparttable}
\end{table}
\vspace{3 mm}

Optical performances have been evaluated at the center of the field of view. However, we also need to figure out the performance over a full field of view since the LAF off-axis system has advantages especially in wide field of view systems \citep{chang2013}. 

\begin{figure}[!ht]
\centering\includegraphics[width=10cm]{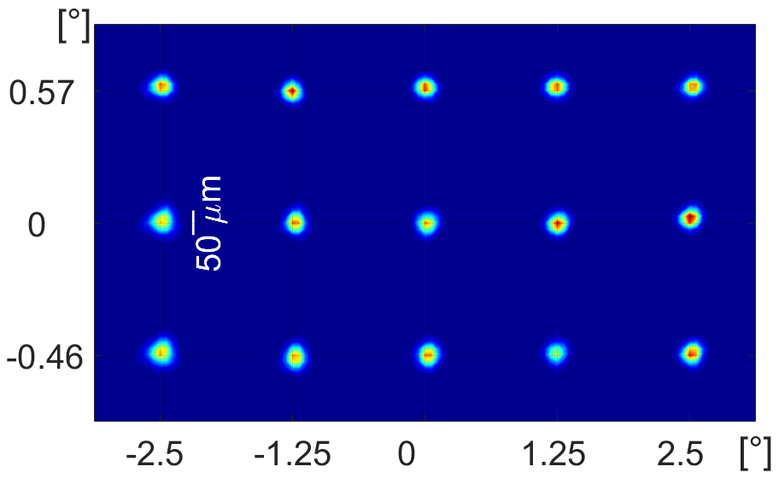}
\caption{\label{fullfield}Full field imaging results with the 100 $\mu$m pinhole in IR2. The white scale bar indicates 50 $\mu$m.}
\end{figure}

The whole limb instrument was rotated with a rotational axis at the center of the M1 surface for the full field of view tests. Figure~\ref{fullfield} illustrates full field of view test results in IR2. As we can clearly see in the figure, there are no dominant off-axis aberrations detected over a full field of view. The geometry of the spot images is almost uniform while optical performance degradations are not noticeable, indicating that field curvature is negligible. As all channels share the same mirror system, off-axis aberrations in other channels will show a pattern similar to that of IR2.

\section{Discussion and Summary}
\label{sec:discussion}
The limb telescope is the main optical system of the MATS satellite. In this study we performed sensitivity analysis and Monte-Carlo simulations to estimate the expected stability and performance of the telescope in space. Total system throughput and relative pointing are characterized for the multi-channel imaging system. Optical performance is evaluated with the flight instrument not only at the image center but also over full field of view.

System tolerance limits were decided by sensitivity analysis and iterative method using Monte-Carlo simulation. From the sensitivity analysis, we conclude that $\alpha$- and $\beta$- tilt of M3 are suitable compensators for alignment errors. Monte-Carlo simulation results indicate the instrument has a 96$\%$ cumulative probability of meeting its required 0.3 MTF. The simulation does not include surface RMS errors of mirrors. From the sensitivity analysis results, however, the surface errors of fabricated flight mirrors, which are 0.034 - 0.062 $\mu$m, barely degrade the optical resolution.

Total system throughput is calculated by measuring reflectivity and transmittance of all optical components. Mirror reflectivity in UV wavelength is 3 - 9 $\%$ lower than our expectations, a discrepancy that likely resulted from scattering from the diamond turned mirror surface. Relative pointing measurements imply that despite the large vertical offset of UV2 (due to filter misalignment), all IR and UV CCDs can simultaneously cover their targets.

For the distortion measurements, a distortion lager than distortion by design is detected in the system. This additional distortion is caused by combinations of mirror surface errors and filter bending by assembly stress. However, thanks to our characterization results, these distortions can be corrected in the final scientific product.

We also measured the optical resolution and found that measured MTF performance is 0.25 MTF at 20 lp/mm for UV2 and 0.25 - 0.54 MTF at 10 lp/mm for IR channels. The results from UV1 are highly uncertain with a measured 80$\%$ EED spot size of 57.81 $\mu$m, which is less than the required resolution. However, these may be due to the low signal to noise ratio from the low power of the collimator lamp at 270 nm wavelength.

Linear astigmatism and field curvature are not detected in the full field of view image. Since the optical design has been optimized to reduce third order aberrations, we have almost uniform optical performance over large field of view. 

In summation, we performed flight model characterizations and performance tests of the MATS satellite. The satellite applies the linear-astigmatism-free confocal off-axis reflective optical design for the first time. It proved that building LAF off-axis system is feasible with the satellite-level optical performance. Optical design and testing method introduced by this paper are applicable to any other optical applications.

\medskip
This work was supported by Swedish National Space Agency (SNSA), Creative Convergence Research Project in the National Research Council of Science and Technology of Korea (CAP–15–01–KBSI). This work was a collaborative research work with the Wyant College of Optical Sciences for their assistance with the linear astigmatism free - three mirror system development. This research was made possible in part by the Technology Research Initiative Fund Optics/Imaging Program at the University of Arizona.

\newcommand{\newblock}{}
\bibliographystyle{aasjournal}
\bibliography{ref_mats}

\end{document}